\documentclass[aps,pra,showpacs,superscriptaddress]{revtex4}
\usepackage{eurosym}
\usepackage{eurosym}
\usepackage{amssymb}
\usepackage{graphicx}
\usepackage{hyperref}

\begin{document}

\title{Bath-induced correlations in an infinite-dimensional Hilbert space}
\author{Marco Nizama$^{1,2}$ and\ Manuel O. C\'{a}ceres}
\email{marco.nizama@fain.uncoma.edu.ar}
\email{caceres@cab.cnea.gov.ar}
\affiliation{Centro At\'{o}mico Bariloche, CNEA, Instituto Balseiro and CONICET, 8400
Bariloche, Argentina.\\
$^{2}$Departamento de F\'{\i}sica, Universidad Nacional del Comahue, 8300,
Neuquen, Argentina}
\date{}

\begin{abstract}
\ \newline
Quantum correlations between two free spinless dissipative distinguishable
particles (interacting with a thermal bath) are studied analytically using\
the quantum master equation and tools of quantum information. Bath-induced
coherence and correlations in an \textit{infinite-dimensional} Hilbert space
are shown. We show that for temperature $T>0$ the time-evolution of the
reduced density matrix cannot be written as the direct product of two
independent particles. We have found a time-scale that characterizes the
time when the bath-induced coherence is maximum before being wiped out by
dissipation (purity, relative entropy, spatial dispersion, and mirror
correlations are studied). The Wigner function associated to the Wannier
lattice (where the dissipative quantum walks move) is studied as an indirect
measure of the induced correlations among particles. We have supported the
quantum character of the correlations by analyzing the geometric quantum
discord.
\end{abstract}

\maketitle

\section{Introduction}

In many practical situations where classical mechanics is enough to make a
good description of a system, the interaction with a surrounding (bath)
leads to dissipation and fluctuations. This program has also been extended
to quantum mechanics concluding with fundamental results which can be
summarized in the Fluctuation-Dissipation theorem, see \cite{Reichl,holanda}
and references therein. Nevertheless, if we wish to describe the quantum
non-equilibrium evolution the problem is inevitably outside of the scope of
the previous Kubo-like approach. Other approximations must be introduced to
work out an \textit{open} quantum mechanics system \cite{Alicki}. Of
importance is the analysis of the quantum mechanics correlations generated
during the elapse of time of the interaction with a measurement apparatus 
\cite{holanda}. In particular quantum mechanics correlations in a bipartite
system have generated much interest for various tasks such as computing \cite%
{nielsen}, imaging and metrology \cite{Kolobov}. Thus, the understanding of
the mechanism of decoherence is an issue of great interest as it would allow
progress in construction of quantum mechanics devices \cite{holanda}.

The effect of a thermal quantum bath $\mathcal{B}$ on a microscopic system $%
\mathcal{S}$ has in particular been much discussed, the general consensus
being that $\mathcal{B}$ leads to dissipation and decoherence on $\mathcal{S}
$. Breaking the isolation of $\mathcal{S}$ is then believed to significantly
increase the decoherence \cite{Schlosshauer,Kendon,nielsen}. Nevertheless,
if this were true for all quantum mechanics\ open systems, no matter how
small is the interaction with $\mathcal{B}$, fluctuation and dissipative
effects would become very costly for the operation of quantum mechanics
devices as needed in quantum computation. Indeed very recently it has been
shown that entanglement between two qubits can be generated if the two
qubits interact with a common thermal bath \cite{Braun}, also research on
quantum information processing --in finite dimensional systems-- have led to
the picture of entanglement as a precious resource \cite{Finite}. Additional
studies concerning the analysis of a common bath vs individual baths have
lead to support the idea of bath-induced correlations in Markovian and
non-Markovian approximations \cite{CommBath}. A related result has also been
found where the role of non-Markovian effects for the quantum entanglement
has been studied \cite{ChemPhysLett}.

In this context an important point of view is achieved if we could analyze
systems associated with an \textit{infinite} dimensional Hilbert space.
However, this is not a simple task for dissipative systems using the quantum
information theory. In this work we propose to study --analytically--
quantum correlations between two particles (in an infinite \textit{discrete}
dimensional Hilbert space) interacting with a thermal bath. Then we will
show that indeed the bath $\mathcal{B}$ generates not only dissipation, but
induces coherence and correlations between particles immersed on it. In
order to prove this fact, we will do exact calculations of the dynamics of
spinless quantum walks \cite{aharanov,vK95,blumen2,manuel}. Then exact
analytic results for the induced correlations can be computed showing that $%
\mathcal{B}$ may generate correlations between particles originally
uncorrelated. In what follows, we present calculations to measure
correlations and so to define a characteristic time-scale for a maximum
coherence in the system before being wiped\ out by dissipation. We prove
that these correlations are of quantum mechanics\ nature using the tools of
quantum information theory, thus we show that our results can be used to
measure the quantum to classical transition as we have already done studying
associated qubit systems \cite{physA-1,physA-2}.

The simplest implementation that reflects the role of a coherent
superposition can be proposed in the framework of quantum walk experiments,
or its numerical simulations \cite{zahringer,schreiber,broome,schmitz,karski}%
. A Dissipative Quantum Walk (DQW) has also been defined as a spinless
particle moving in a lattice and interacting with a phonon bath \cite%
{aharanov,vK95,kempe,manuel,mm,physA-1,physA-2}. These models can also be
extended considering a many-body description as we present here. In
particular in this work we will implement explicit calculations for a system
constituted by two distinguishable DQW's. Therefore, our approach can be
used to tackle this general problem pointing out the interplay between
dissipation and the bath-particle interaction.

A non-Markov extension of the present DQW model can also be introduced using
the Continuous-time Random Walk (CTRW) approach introduced by Montroll and
Weiss \cite{Montroll,libro}. In appendix A a short review on the subject of
quantum jumps in presented, this picture can also be related to the random
quantum maps approach in the context of the Renewal theory \cite{caceres2017}%
.

\section{Dissipative Quantum Walks}

A model of two free distinguishable particles coupled to a common bath $%
\mathcal{B}$ can be written using the Wannier base in the following way. Let
the total Hamiltonian be $H_{\mathcal{T}}=H_{\mathcal{S}}+H_{\mathcal{B}}+H_{%
\mathcal{SB}}$, here $H_{\mathcal{S}}$ is the \textit{free} \textit{%
tight-binding} Hamiltonian \cite{Two-QW} (our system $\mathcal{S}$), which
can be written in the form: 
\[
H_{\mathcal{S}}=2E_{0}\mathbf{I}-\frac{\Omega }{2}\left( a_{12}^{\dag
}+a_{12}\right) , 
\]%
here $\{a_{12}^{\dag },a_{12}\}$ are shift operators for the particles
labeled $1$ and $2$, and $\mathbf{I}$ is the identity in the Wannier base%
\begin{equation}
\mathbf{I}=\sum_{s,s^{^{\prime }}}\arrowvert s,s^{^{\prime }}\rangle \langle
s,s^{^{\prime }}\arrowvert,
\end{equation}%
then:%
\begin{equation}
a_{12}^{\dag }\arrowvert s_{j},s_{l}\rangle =\arrowvert s_{j}+1,s_{l}\rangle
+\arrowvert s_{j},s_{l}+1\rangle  \label{RR1}
\end{equation}%
\begin{equation}
a_{12}\arrowvert s_{j},s_{l}\rangle =\arrowvert s_{j}-1,s_{l}\rangle +%
\arrowvert s_{j},s_{l}-1\rangle .  \label{RR2}
\end{equation}%
We note that a \textquotedblleft shift operator\textquotedblright\
translates each particle individually. Here we have used a \textquotedblleft
pair-ordered\textquotedblright\ brac-ket $\arrowvert s_{j},s_{l}\rangle $
representing the particle \textquotedblleft $1$\textquotedblright\ at site $%
s_{j}$ and particle \textquotedblleft $2$\textquotedblright\ at site $s_{l}$%
. From Eqs. (\ref{RR1})-(\ref{RR2}) it is simple to see that $\left[
a_{12}^{\dag },a_{12}\right] =0$ and the fact that 
\[
a_{12}a_{12}^{\dag }\arrowvert s_{j},s_{l}\rangle =2\arrowvert %
s_{j},s_{l}\rangle +\arrowvert s_{j}-1,s_{l}+1\rangle +\arrowvert %
s_{j}+1,s_{l}-1\rangle . 
\]

%
%
%
%
%
%
%
%
%
%
%
%
%
%
%
%
%
$H_{\mathcal{B}}$ is the phonon bath Hamiltonian $H_{\mathcal{B}%
}=\sum\limits_{n}\hbar \omega _{n}\mathcal{B}_{n}^{^{\dag }}\mathcal{B}_{n},$
thus $\{\mathcal{B}_{n}^{^{\dag }},\mathcal{B}_{n}\}$ are bosonic operator
characterizing the thermal bath at equilibrium.

The term $H_{\mathcal{SB}}$ in the total Hamiltonian represents the
interaction term between $\mathcal{S}$ and $\mathcal{B}$, here we use a
linear interaction between the particles and the bath operators. Our model
is a many-body generalization of the van Kampen approach used to address the
nature of a physical dissipative particle interacting with a boson bath \cite%
{vK95}. Because the shift operators $a_{1,2}$ and $a_{12}^{\dag }$ are
constant of motions, any bath interaction with these shift operators will
lead to a completely positive infinitesimal generator, see Kossakowski and
Lindblad \cite{Alicki}. Thus, for two \textit{distinguishable} particles we
propose the interaction term $H_{\mathcal{SB}}$ in the form%
\begin{equation}
H_{\mathcal{SB}}=\hbar \Gamma \left( a_{12}\otimes \sum\limits_{n}v_{n}%
\mathcal{B}_{n}+a_{12}^{\dag }\otimes \sum\limits_{n}v_{n}^{\ast }\mathcal{B}%
_{n}^{^{\dag }}\right) ,  \label{Hi}
\end{equation}%
where $v_{n}$ represents the spectral intensity weight function from the
phonon bath at thermal equilibrium, and $\Gamma $ is the interaction
parameter in the model. We have chosen this interaction Hamiltonian in order
to recover the classical master equation for two independent random walk in
the case when $\Omega =0$, for a more extended discussion on this issue see
Appendix A in \cite{physA-1,libro}.

In order to study the non-equilibrium evolution of $\mathcal{S}$ we derive
from $H_{\mathcal{T}}$, eliminating the bath variables, a dissipative
quantum infinitesimal generator (see appendix A). Tracing out bath variables
in the Ohmic approximation and assuming as initial state of the total system
a density matrix in the form of a product 
\[
\rho _{T}\left( 0\right) =\rho \left( 0\right) \otimes e^{-H_{\mathcal{B}%
}/k_{B}T}/Z, 
\]%
where $Z=$Tr$\left( e^{-H_{\mathcal{B}}/k_{B}T}\right) $, we can write the
Markov Quantum Master Equation (QME) \cite{Alicki,holanda,manuel}:

\begin{eqnarray}
\dot{\rho} &=&\frac{-i}{\hbar }\left[ H_{eff},\rho \right] +\frac{D}{2}%
\left( 2a_{12}\rho a_{12}^{\dag }-a_{12}^{\dag }a_{12}\rho -\rho
a_{12}a_{12}^{\dag }\right)  \nonumber \\
&+&\frac{D}{2}\left( 2a_{12}^{\dag }\rho a_{12}-a_{12}a_{12}^{\dag }\rho
-\rho a_{12}^{\dag }a_{12}\right) ,  \label{QME}
\end{eqnarray}%
here $D\equiv \Gamma ^{2}k_{B}T/\hbar $, where $T$ is the temperature of the
bath $\mathcal{B}$. In the present paper we are interested in solving this
QME with a localized initial condition in the Wannier lattice, i.e.,: $\rho
(0)=\arrowvert s_{1},s_{2}\rangle \langle s_{1},s_{2}\arrowvert.$

Adding $-2E_{0}+\Omega $ to $H_{\mathcal{T}}$ the effective Hamiltonian
turns to be%
\begin{equation}
H_{eff}=\Omega \left( \mathbf{I}-\frac{a_{12}^{\dag }+a_{12}}{2}\right)
-\hbar \omega _{c}a_{12}a_{12}^{\dag },  \label{Heff}
\end{equation}%
where $\omega _{c}\equiv 2\tilde{\omega}_{c}\Gamma ^{2}$ is related to the
frequency cut-off $\tilde{\omega}_{c}$ in the Ohmic approximation \cite%
{vK95,Alicki,manuel}. It can be seen from the \textit{strength function }$%
g\left( \omega \right) $ of thermal oscillators (defined by $g\left( \omega
\right) \Delta \omega \leftrightarrow \left[ \sum_{n}v_{n}^{2}\right]
_{\{\omega <\omega _{n}<\omega +\Delta \omega \}}$), that the high-frequency
oscillators (beyond $\tilde{\omega}_{c}$) only modify the effective
Hamiltonian, that is its unitary evolution, see appendix A. This von Neumann
dynamics can be defused by going to the interaction representation. However,
here we will be interested in studying the non-equilibrium evolution of the
system as a function of the rate of energies $\Omega $ and $D\equiv \Gamma
^{2}k_{B}T/\hbar $. Then, in order to simplify the analysis of the QME (\ref%
{QME}) we will drop-out the term $\hbar \omega _{c}a_{12}a_{12}^{\dag }$ in
the effective Hamiltonian, which only produces additional reversible
coherence. Under the assumption that $\hbar \tilde{\omega}_{c}/k_{B}T\ll 1$,
the dissipative coefficient $D$ appearing in (\ref{QME}) depends on the
strength function and the thermal bath correlation function. This terms only
involves oscillators in the low-frequency region. It is also possible to see
that the Markov approximation used to get (\ref{QME}) involves a \textit{%
coarse-grained} time scale such that $\tilde{\omega}_{c}\Delta t\gg 1$ in
addition to a second order weak interaction approach, which at the end leads
to the damping dissipative factor $D$.

From (\ref{QME}) is simple to see that as temperature vanishes ($%
D\rightarrow 0$) the unitary evolution is recovered. On the contrary, the
case $\Omega \rightarrow 0$ (or $D\rightarrow \infty $ ) would correspond to 
\textit{two} classical random walks. We note, however, that for the present
two-body \textit{quantum} problem, when $D/\hbar \Omega \gg 1$, the
classical profile cannot be recovered because correlations between particles
are induced from thermal bath $\mathcal{B}$. In addition, here we note that
the initial condition of particles would be relevant for the calculation of
the time-dependent bath-induced decoherence.

\subsection{Solution for two QDW}

We will solve this QME (\ref{QME}) with a localized initial condition in the
Wannier lattice, i.e.,:%
\begin{equation}
\rho (0)=\arrowvert s_{1},s_{2}\rangle \langle s_{1},s_{2}\arrowvert=%
\arrowvert\vec{0}\rangle \langle \vec{0}\arrowvert.  \label{IC}
\end{equation}%
The operational calculus in the QME\ will be done using Wannier vector
states to evaluate elements of the density matrix $\rho (t)$.

To solve Eq.(\ref{QME}) we apply $\arrowvert s_{1},s_{2}\rangle $ from the
right and $\langle s_{1},s_{2}\arrowvert$ from the left, then using Eqs.(\ref%
{RR1}) and (\ref{RR2}) the evolution equation can be written in terms of the
usual Wannier \textit{"brac-ket"}. Therefore, we can introduce the discrete
Fourier transform noting that a Fourier "\textit{brac-ket}" is defined in
terms of a Wannier basis for \textit{two particles} in the form:%
\[
\left\vert k_{1},k_{2}\right\rangle =\frac{1}{2\pi }\sum\limits_{s_{1},s_{2}%
\in \mathcal{Z}}e^{ik_{1}s_{1}}e^{ik_{2}s_{2}}\left\vert
s_{1},s_{2}\right\rangle , 
\]%
with $k_{j}\in (-\pi ,\pi )$ and $s_{1},s_{2}\in $ integers. Thus finally
the QME (\ref{QME}) can be written as:

\[
\left\langle k_{1},k_{2}\left\vert \frac{d\rho }{dt}\right\vert
k_{1}^{\prime },k_{2}^{\prime }\right\rangle \!\!=\mathcal{F}%
(k_{1},k_{1}^{\prime },k_{2},k_{2}^{\prime })\ \!\!\left\langle
k_{1},k_{2}\left\vert \rho \right\vert k_{1}^{\prime },k_{2}^{\prime
}\right\rangle , 
\]%
here%
\[
\mathcal{F}(k_{1},k_{1}^{\prime },k_{2},k_{2}^{\prime })\equiv \{\mathcal{F}%
^{(1)}(k_{1},k_{1}^{\prime })+\mathcal{F}^{(1)}(k_{2},k_{2}^{\prime })+2D[%
\mathbf{C}\left( k_{1}\!,k_{2}^{\prime }\right) +\mathbf{C}\left(
k_{2}\!,k_{1}^{\prime }\right) -\mathbf{C}\left( k_{1}\!,k_{2}\right) -%
\mathbf{C}\left( k_{1}^{\prime }\!,k_{2}^{\prime }\right) ]\}, 
\]%
where%
\[
\mathcal{F}^{(1)}(k_{2},k_{2}^{\prime })\equiv \left[ \!\frac{-i}{\hbar }%
\!\left( \mathcal{E}_{k_{2}}\!-\!\mathcal{E}_{k_{2}^{\prime }}\right)
\!+\!2D\!\left( \mathbf{C}\left( k_{2}\!,k_{2}^{\prime }\right) -\!1\right)
\!\right] , 
\]%
is the one-particle infinitesimal generator in the Fourier representation 
\cite{mm},%
\[
\mathbf{C}\left( k_{1}\!,k_{2}\right) \equiv \cos \left(
k_{1}\!-\!k_{2}\right) \!\text{ and }\mathcal{E}_{k_{i}}\equiv \Omega
\left\{ 1-\cos k_{i}\right\} . 
\]%
Note that $\mathcal{F}(k_{1},k_{1},k_{2},k_{2})=0$ leading to a
momentum-like conservation law: $\langle k_{1},k_{2}|\frac{d\rho (t)}{dt}%
|k_{1},k_{2}\rangle =0$.

Elements of $\rho (t)$ can be calculated in the Wannier basis 
\[
\arrowvert s_{1},s_{2}\rangle =\frac{1}{2\pi }\int_{-\pi }^{\pi }\int_{-\pi
}^{\pi }dk_{1}dk_{2}\ e^{-ik_{1}s_{1}}e^{-ik_{2}s_{2}}\arrowvert %
k_{1},k_{2}\rangle . 
\]%
After some algebra and using Bessel's function properties we can write an
analytical formula for $\rho (t)$ in Wannier representation $\langle
s_{1},s_{2}|\rho (t)|s_{1,}^{\prime }s_{2}^{\prime }\rangle $ (to simplify
the notation we use $t_{\Omega }\equiv \frac{\Omega t}{\hbar },t_{D}\equiv
2Dt$ whenever it is necessary
)

\begin{eqnarray}
\langle s_{1},s_{2}|\rho (t)|s_{1}^{\prime },s_{2}^{\prime }\rangle \!\!\!
&=&\!\!i^{(s_{1}-s_{1}^{\prime }+s_{2}-s_{2}^{\prime
})}e^{-2t_{D}}\!\!\!\!\!\!\!\!\!\!\!\!\!\!\!\!\!\!\!\!\sum_{%
\{n_{1},n_{2},n_{3},n_{4},n_{5},n_{6}\}\in \mathcal{Z}}\!\!\!\!\!\!\!\!\!\!%
\!\!\!\!\!\!(-1)^{n_{4}+n_{5}}  \label{rhog} \\
&\times &\!\!J_{s_{1}+n_{1}+n_{2}+n_{5}}\!\!\left( t_{\Omega }\right) \!\!\
J_{s_{1}^{\prime }+n_{1}+n_{3}+n_{4}}\!\!\left( t_{\Omega }\right)  \nonumber
\\
&\times &\!\!J_{s_{2}+n_{3}-n_{5}+n_{6}}\!\!\left( t_{\Omega }\right) \!\!\
J_{s_{2}^{\prime }+n_{2}-n_{4}+n_{6}}\!\!\left( t_{\Omega }\right)
\!\!\prod\limits_{n_{i}=1}^{6}I_{n_{i}}\!\left( t_{D}\right) ,\
\{s_{j},s_{l}^{\prime }\}\in \mathcal{Z}  \nonumber
\end{eqnarray}%
where $J_{n}$ and $I_{n}$ are Bessel's functions of integer order $n\in 
\mathcal{Z}$. These functions satisfy that%
\[
J_{-n}(t)=(-1)^{n}J_{n}(t),\ J_{n}(-t)=(-1)^{n}J_{n}(t), 
\]%
and 
\[
I_{-n}(t)=I_{n}(t),\ I_{n}(-t)=(-1)^{n}I_{n}(t). 
\]%
%
%
%
%
%
%
%
%
%
%
%
%
%
%
%
This solution is symmetric under the exchange of particles \cite{particles}
(therefore preserving the symmetry of the initial condition), is Hermitian
and fulfills normalization in the lattice $\mbox{Tr}\lbrack \rho
(t)]=\sum_{\{s_{1},s_{2}\}\in \mathcal{Z}}\langle s_{1},s_{2}|\rho
(t)|s_{1},s_{2}\rangle =1$, $\forall t$; positivity is assured because the
infinitesimal generator fulfills the structural theorem \cite{Alicki}. The
probability of finding one particle in site $s_{1}$ and another in $s_{2}$
is given by the probability profile: $P_{s_{1},s_{2}}(t)\equiv \langle
s_{1},s_{2}|\rho (t)|s_{1},s_{2}\rangle $ and shows the expected reflection
symmetry in the plane: $s_{1}-s_{2}=0$.

In the case $D=0$, i.e., a quantum closed system without dissipation, we
recover the solution for two quantum walk: $\langle s_{1},s_{2}|\rho
(t)|s_{1}^{\prime },s_{2}^{\prime }\rangle
_{D=0}=\prod_{j=1}^{2}i^{(s_{j}-s_{j}^{\prime })}J_{s_{j}}\!\left( t_{\Omega
}\right) \!J_{s_{j}^{\prime }}\!\left( t_{\Omega }\right) $, this means that
from an uncorrelated initial condition $\rho (0)$, the solution $\rho (t\geq
0)_{D=0}$ is written as the direct product of two independent quantum
particles.

As we mentioned before a \textit{classical} random walk regime \cite%
{Reichl,libro} cannot be recovered. For $D\gg \Omega /\hbar $ the two-body
density matrix is $\rho (t)\neq \rho _{1}(t)\otimes \rho _{2}(t)$, showing a
complex pattern structure in terms of convolutions of classical profiles.
From Eq.(\ref{rhog}) it can be proved that when $D\gg \Omega /\hbar $ we get%
\[
\lim_{D\gg \Omega /\hbar }P_{s_{1},s_{2}}(t)\neq P_{s_{1}}(t)\times
P_{s_{2}}(t)=e^{-2t_{D}}\ I_{s_{1}}\left( t_{D}\right) I_{s_{2}}\left(
t_{D}\right) , 
\]%
here $P_{s_{j}}$ is the classical probability profile for each particle. So
a classical regime [for $t\rightarrow \infty $] cannot be obtained. This
means that the profile for two DQW will not be a Gaussian bell-shape in 2D.
In addition, it is intriguing to note that from the QME there exist an
important competition between building correlations vs inducing dissipative
decoherence.

Note that the one-particle density matrix is recovered tracing-out the
degrees of freedom of the second one, say $j=2$: 
\[
\rho ^{(1)}(t)\equiv \mbox{Tr}_{2}[\rho (t)], 
\]%
then%
\[
\langle s_{1}|\rho ^{(1)}(t)|s_{1}^{\prime }\rangle =i^{(s_{1}-s_{1}^{\prime
})}e^{-t_{D}}\!\!\sum_{n\in \mathcal{Z}}\!\!\ J_{s_{1}+n}\!\!\left(
\!t_{\Omega }\right) \!\!\ J_{s_{1}^{\prime }+n}\!\!\left( \!t_{\Omega
}\right) \!\!\ I_{n}\!\left( t_{D}\right) , 
\]%
solution that indeed shows, for $D\gg \Omega /\hbar $, a random walk
behavior for one particle \cite{mm}.\medskip\ In addition we note that in
the lattice the classical random walk solution is $P_{t}(s)=e^{-2Dt}\
I_{s}\left( 2Dt\right) $, and from this expression it is simple to get the
Gaussian profile in the continuous limit \cite{libro}.

\section{Correlations and coherence in the Infinite Dimension Hilbert Space}

\subsection{Purity}

To measure the influence from $\mathcal{B}$ into $\mathcal{S}$ we study the
Purity $\mathcal{P}_{Q}^{(2)}(t)\equiv \mbox{Tr}\lbrack \rho (t)^{2}]$ \cite%
{nielsen}.

\begin{eqnarray}
\mathcal{P}_{Q}^{(2)}(t) &=&\mbox{Tr}\lbrack \rho
(t)^{2}]=\sum_{s_{1},s_{2}=-\infty }^{\infty }\sum_{s_{1}^{\prime
},s_{2}^{\prime }=-\infty }^{\infty }\langle s_{1},s_{2}|\rho
(t)|s_{1}^{\prime },s_{2}^{\prime }\rangle \langle s_{1}^{\prime
},s_{2}^{\prime }|\rho (t)|s_{1},s_{2}\rangle  \nonumber \\
&=&e^{-8Dt}\sum_{m=-\infty }^{\infty }I_{m}\left( 4Dt\right) \sum_{\alpha
,\beta =-\infty }^{\infty }(-1)^{\alpha +\beta }I_{\alpha }\left( 4Dt\right)
I_{\beta }\left( 4Dt\right) I_{\alpha +m}\left( 4Dt\right) I_{\beta
+m}\left( 4Dt\right) I_{\alpha +\beta +m}\left( 4Dt\right) .
\end{eqnarray}%
We can prove that $\mathcal{P}_{Q}^{(2)}(t\geq 0)=1$ for $D=0$, and for $%
D\neq 0$ we get $\mathcal{P}_{Q}^{(2)}(t)\leq 1$ decreasing in the course of
time. Interestingly, for $D\neq 0$, $\mathcal{P}_{Q}^{(2)}(t)$ is different
from the Purity for \textit{two particles} with independent quantum baths,
i.e., $\mathcal{P}_{Q}^{(2)}(t)\neq \mathcal{P}_{Q}^{(1)}(t)\mathcal{P}%
_{Q}^{(1)}(t)$, where 
\[
\mathcal{P}_{Q}^{(1)}(t)=e^{-4Dt}I_{0}\left( 4Dt\right) , 
\]%
is the one-particle Purity (with independent bath \cite{mm}). Thus a common
bath produces a difference in the total purity 
\[
\Delta \mathcal{P}_{Q}\equiv \mathcal{P}_{Q}^{(2)}(t)-\mathcal{P}%
_{Q}^{(1)}(t)\mathcal{P}_{Q}^{(1)}(t)\geq 0, 
\]%
which shows the occurrence of bath-induced correlations.

An outstanding conclusion can be observed by introducing a \textit{change of
basis} in the representation of the two-particle density matrix $\rho
(t)\equiv \rho (\Omega ,D,t)$. Using the time-dependent unitary
transformation:%
\[
\langle s_{1},s_{2}\arrowvert U_{1}\arrowvert s_{1}^{\prime },s_{2}^{\prime
}\rangle =i^{\left( s_{1}+s_{2}+s_{1}^{\prime }+s_{2}^{\prime }\right)
}J_{s_{1}-s_{1}^{\prime }}(\!\frac{\Omega t}{\hbar })J_{s_{2}-s_{2}^{\prime
}}(\!\frac{\Omega t}{\hbar }) 
\]%
in Eq.(\ref{rhog}) it is possible to prove that 
\[
U_{1}\rho (\Omega ,D,t)U_{1}^{\dag }=\rho (\Omega =0,D,t). 
\]%
Thus, properties as Purity $\mathcal{P}_{Q}^{(2)}(t)$, Information Entropy $%
S(t)=-Tr\left[ \rho \ln \rho \right] $ (von Neumann's entropy) can
straightforwardly be shown in this new representation, see Fig.\ref%
{fig-puri-entro}(a),(b) with and without a common bath. As $\rho _{\mathcal{S%
}+\mathcal{B}}(0)$ is a pure state, $S(t)$ is a good measure for the
entanglement between the two particles with $\mathcal{B}$. We noted that
even when the Purity is related to the Information Entropy, $\mathcal{P}%
_{Q}^{(2)}$ gives much insights: we see that $\rho (t)$ for two DQW's with a 
\textit{common} bath the system has more Purity than the case of two DQW's
with \textit{independent} baths. The inset Fig.\ref{fig-puri-entro}(b) shows
the difference $\Delta \mathcal{P}_{Q}$ from the mentioned cases showing a
maximum of correlation for $t_{D}^{\max }\approx 1.2$ before the dissipation
wipes out the bath-induced coherence.

\subsection{Quantum mirror correlations}

\begin{figure}[t]
\includegraphics[width=7.cm,height=7.5cm]{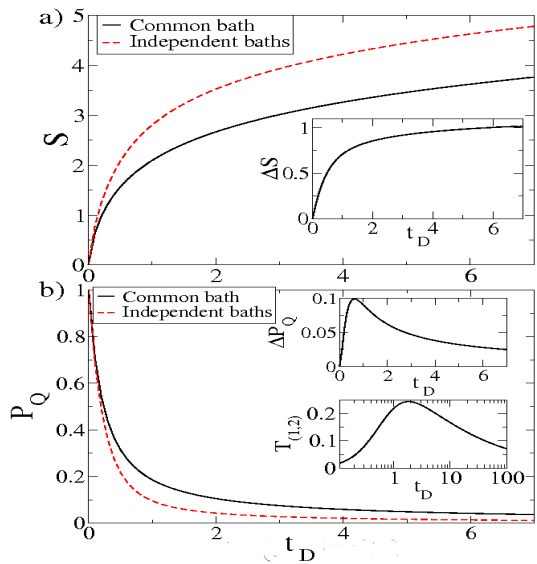}
\caption{(Color online) (a) Information Entropy and (b) Purity for two
different cases (using an initial condition as in Eq.$7$): two DQW's with a
common $\mathcal{B}$ and two DQW's with independent baths. Insets show the
corresponding differences, and $\mathcal{T}_{(1,2)}$; all as a function of $%
t_{D}=2Dt$.}
\label{fig-puri-entro}
\end{figure}

Another measure to quantify the correlations \textit{build up} between the
particles, can be evaluated calculating correlation events for two
particles. We define the total-mirror correlation $\mathcal{T}_{(1,2)}$, $%
\forall \{\Omega ,D\}$, as:%
\[
\mathcal{T}_{(1,2)}=\sum_{s_{1},s_{2}}\langle s_{1},s_{2}\arrowvert\rho (t)%
\arrowvert-s_{1},-s_{2}\rangle -\left( \mathcal{T}^{(1)}\right) ^{2}, 
\]%
where%
\[
\mathcal{T}^{(1)}=\sum_{s_{1},s_{2},s_{2}^{\prime }}\langle s_{1},s_{2}%
\arrowvert\rho (t)\arrowvert-s_{1},s_{2}^{\prime }\rangle
=e^{-2Dt}I_{0}(2Dt). 
\]%
The quantity $\mathcal{T}^{(1)}$ can be interpreted as the \textit{%
one-particle} classical random walk return to the origin \cite{libro}, to
see this note that%
\[
\rho ^{(1)}(t)=\mbox{Tr}_{2}[\rho (t)], 
\]%
and so%
\[
\sum_{s_{1}}\langle s_{1}\arrowvert\rho ^{(1)}(t)\arrowvert-s_{1}\rangle
=e^{-2Dt}I_{0}(2Dt). 
\]%
In the inset of Fig.\ref{fig-puri-entro}(b) we plot the correlation $%
\mathcal{T}_{(1,2)}(t)$ showing that there is a time-scale when this quantum
correlation reaches a maximum $t_{D}^{\max }\equiv 2Dt^{\max }\simeq
1.9\cdots $ before the long-time asymptotic regime $\sim 1/t$,
characterizing the decoherence in the two-particles system.

\subsection{Relative entropy of coherence}

In this section, we will quantify the quantum coherence in our system. For
such purpose we use a entropic measure of the quantum coherence called the
Relative Entropy of Coherence \cite{Plenio2014,mauro}.

\begin{figure}[t]
\includegraphics[width=8.5cm,height=6.5cm]{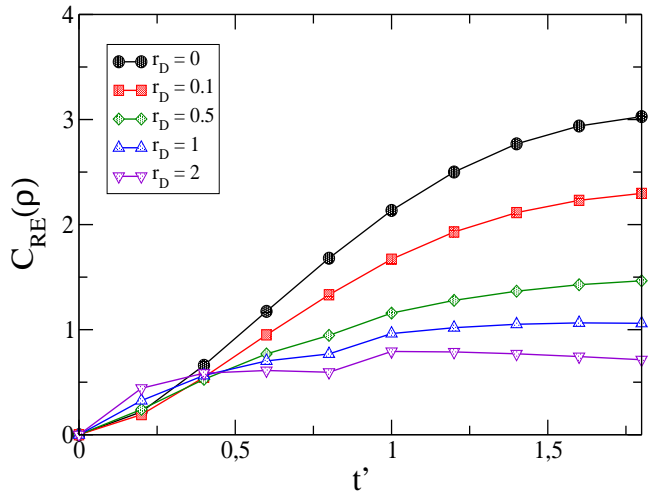}
\caption{(Color online) Relative entropy of coherence for two particles with
localized initial condition (see Eq.\protect\ref{IC}) as function of $%
t^{\prime }=t_{\Omega }\equiv \Omega t/\hbar $. This function shows a
crossover at $t^{\prime }\simeq 0.4$ as a function of time $t^{\prime }$.
The $C_{RE}$ is calculated for different values of the dissipation parameter 
$r_{D}=\frac{2D}{\Omega /\hbar }.$}
\label{fig-ERC}
\end{figure}

For any quantum state $\rho =\rho \left( t\right) $ on the Hilbert space $%
\mathcal{H}$, the Relative Entropy of Coherence is defined as 
\[
C_{RE}=\emph{S}\left( \rho _{diagnal}\right) -\emph{S}\left( \rho \right) , 
\]%
where $\emph{S}\left( \rho \right) =Tr\left[ \rho \ln \rho \right] $ is the
von Neumann entropy. In \cite{Plenio2014}, Baumgratz et al. shown that the
Relative Entropy of Coherence is a good measure of the quantum coherence. In
what follows we will use Wannier's basis to calculate $C_{RE}$. The results
of this measure is shown in figure \ref{fig-ERC}. We have studied the $%
C_{RE} $ as a function of $t^{\prime }=t_{\Omega }=\frac{\Omega t}{\hbar }$
and we use several values of rescaled dissipation parameter $r_{D}\equiv 
\frac{2D}{\Omega /\hbar }=0,$ $0.1,$ $0.5,$ $1,$ $2$. These results can also
be used as an indicator that bath has created correlations between particles
for $t^{\prime }<\tau _{c}\approx 0.4$. For long times $t^{\prime }>\tau
_{c} $ the function $C_{RE}$ decreases with increasing $r_{D}$; this means
that there is a strong competition between the \textit{bath-induced coherence%
} and the \textit{inherent} \textit{decoherence }due to dissipation.
Long-time values of $C_{RE}$ are not plotted due to numerical computer
limitations.

\subsection{Quantum profile coherence}

\begin{figure}[t]
\includegraphics[width=7.8cm,height=11.cm]{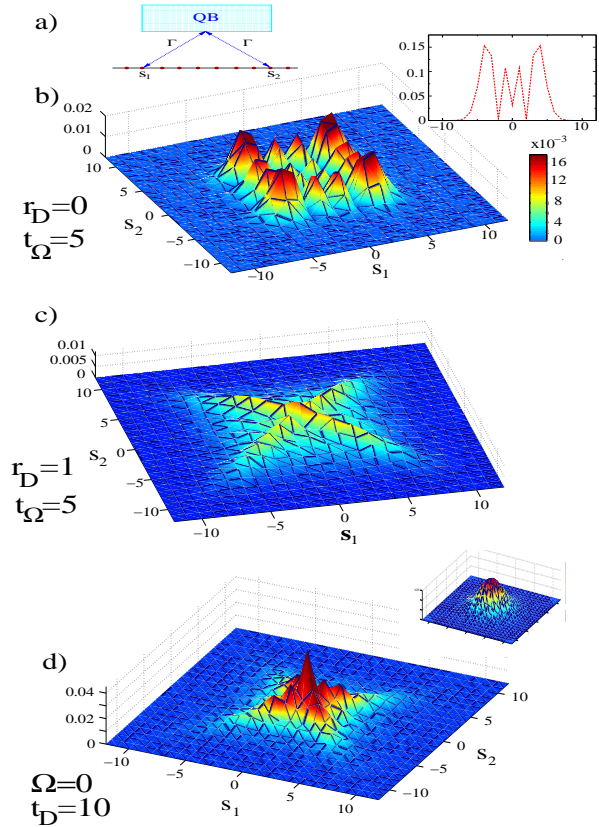}
\caption{(Color online) (a) Representation of two DQW's interacting with $%
\mathcal{B}$. (b) Profile as a function of $s_{1}$ and $s_{2}$. The inset
shows the profile tracing out the second particle. (c) Here the pattern can
be seen even when $D\neq 0$. (d) Profile for $r_{D}\gg 1$ (or $\Omega
\rightarrow 0$). The inset shows the Gaussian bell-shape for two particles
with independent baths. Blue indicates, roughly, the value zero while red a
high value of probability.}
\label{fig-prob2DQW}
\end{figure}

As before, let us use $r_{D}$ the rate of energy scales $r_{D}\equiv \frac{2D%
}{\Omega /\hbar }$ and $t^{\prime }$ a dimensionless time (depending on the
plotting we used $t_{\Omega }$ or $t_{D}$). In Fig.\ref{fig-prob2DQW}(b),
(c), (d) we show the probability profile for having particles at the site $%
s_{1}$ and $s_{2}$, i.e., $P_{s_{1},s_{2}}(t^{\prime }=t_{\Omega })$ for
different values of $r_{D}$ (see Eq.(\ref{rhog})). Here, Fig.\ref%
{fig-prob2DQW}(b) corresponds to the case when the two particles do not
interact with the bath ($D=0$), the inset shown the one-axis projection of
one tight-binding quantum walk \cite{mm}. In Fig.\ref{fig-prob2DQW}(d) $%
P_{s_{1},s_{2}}(t^{\prime }=t_{D})$ corresponds to the high dissipative
regime.

When $D>0$ the probability profile is modified appearing interference
patterns along of line $s_{1}=s_{2}$, raising the value of the probability
in the direction $s_{1}=-s_{2}$ (conservation of total momentum), see Fig.%
\ref{fig-prob2DQW}(c). In the case $\Omega \rightarrow 0$ (or $r_{D}\gg 1$),
the $P_{s_{1},s_{2}}(t^{\prime }=t_{D})$ shows a different pattern signing
the quantum nature in its profile, see Fig.\ref{fig-prob2DQW}(d). This is in
contrast to the case of two particles with independent baths $\mathcal{B}%
_{1} $ and $\mathcal{B}_{2}$ in this case the probability profile is a
Gaussian bell-shape as is shown in the inset. We remark that for two DQW
with a common bath the probability profile can never be represented as a
Gaussian distribution due to the bath-induced correlations.

\subsection{Geometric quantum discord}

To characterize the quantum correlations in the system we use the geometric
quantum discord measure \cite{GQD,lowerGQD,3GQD,Rau}, which is easier to
obtain instead of original quantum discord measure (which involves an
optimization procedure \cite{Discord}), and it has been proved to be a
necessary and sufficient condition for non-zero quantum discord \cite{GQD}.

The geometric quantum discord (GQD) is defined as%
\begin{equation}
D_{G}\left( \rho \right) =\min_{\chi \in \Omega _{0}}\left\vert \left\vert
\rho -\chi \right\vert \right\vert ^{2},  \label{GQD}
\end{equation}%
where $\Omega _{0}$ denotes the set of zero-discord states and $\left\vert
\left\vert X-Y\right\vert \right\vert ^{2}=\mbox{Tr}\left( X-Y\right) ^{2}$
is the square norm in the Hilbert-Schmidt space. Additionally, the lower
bound of the GQD is calculated using the density operator, which is defined
on a bipartite system (belonging to $H^{a}\otimes H^{b}$, with $\dim H^{a}=m$
and $\dim H^{b}=n$) \cite{GQD,lowerGQD,3GQD,Rau} as:%
\begin{eqnarray}
\rho &=&\frac{1}{mn}\left( \mathbf{I}_{m}\otimes \mathbf{I}_{n}+\sum_{i}x_{i}%
\tilde{\lambda}_{i}\otimes \mathbf{I}_{n}+\sum_{j}y_{j}\mathbf{I}_{m}\otimes 
\tilde{\lambda}_{j}\right.  \label{GQD1} \\
&&+\left. \sum_{j}t_{ij}\tilde{\lambda}_{i}\otimes \tilde{\lambda}%
_{j}\right) ,  \nonumber
\end{eqnarray}%
here $\tilde{\lambda}_{i},i=1,\cdots ,m^{2}-1$ and $\tilde{\lambda}%
_{j},j=1,\cdots ,n^{2}-1$ are the generators of $SU(m)$ and $SU(n)$
respectively, satisfying $\mbox{Tr}\left( \tilde{\lambda}_{i}\tilde{\lambda}%
_{j}\right) =2\delta _{ij}$, and $\mathbf{I}_{m}$ is the identity operator
in $m$-dimension. In this expression the vectors $\vec{x}\in \emph{R}%
^{m^{2}-1}$ and $\vec{y}\in \emph{R}^{n^{2}-1}$ of the subsystems $A$ and $B$
are given by: 
\begin{eqnarray*}
x_{i} &=&\frac{m}{2}\mbox{Tr}\left( \rho \tilde{\lambda}_{i}\otimes \mathbf{I%
}_{n}\right) =\frac{m}{2}\mbox{Tr}\left( \rho _{A}\tilde{\lambda}_{i}\right)
\\
y_{j} &=&\frac{m}{2}\mbox{Tr}\left( \rho \mathbf{I}_{m}\otimes \tilde{\lambda%
}_{j}\right) =\frac{n}{2}\mbox{Tr}\left( \rho _{B}\tilde{\lambda}_{j}\right)
,
\end{eqnarray*}%
and the correlation matrix $T\equiv \left[ t_{ij}\right] $ is given by 
\[
T\equiv \left[ t_{ij}\right] =\frac{mn}{4}\mbox{Tr}\left( \rho \tilde{\lambda%
}_{i}\otimes \tilde{\lambda}_{j}\right) . 
\]

The lower bound of the GQD is calculated in the following form: 
\begin{equation}
D_{G}\left( \rho \right) \geq \frac{2}{m^{2}n}\left( \left\vert \left\vert 
\vec{x}\right\vert \right\vert ^{2}+\frac{2}{n}\left\vert \left\vert
T\right\vert \right\vert ^{2}-\sum_{i=1}^{m-1}\eta _{i}\right) ,
\label{GQD2}
\end{equation}%
where $\eta _{i},i=1,2,\cdots ,m^{2}-1$ are eigenvectors of the matrix $%
\left( \vec{x}\vec{x}^{t}+\frac{2}{n}TT^{t}\right) $ arranged in
non-increasing order \cite{lowerGQD}.

\subsubsection{Lattice bipartition, the qubit-qubit set}

We need to define a procedure on the lattice in order to study the GQD (a
similar approach has been done in \cite{physA-2}), in this context
introducing a bipartition we will end with a \textit{qutrit-qutrit system}.

\begin{figure}[t]
\includegraphics[width=8.5cm,height=6.5cm]{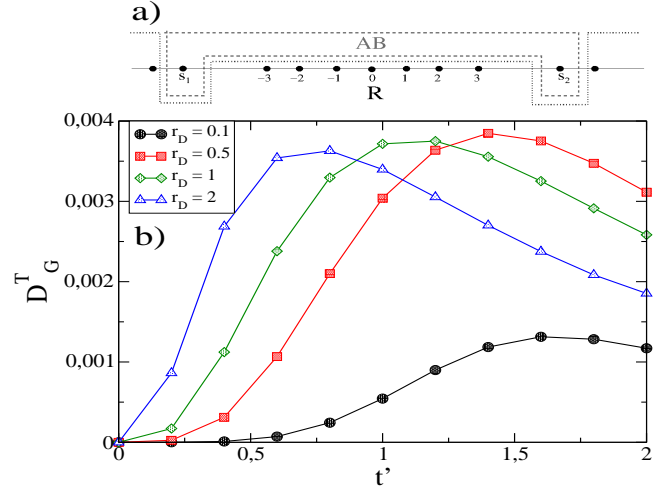}
\caption{(a) The bipartition to calculated the GQD. (b) The function $%
D_{G}^{T}\left( \protect\rho _{AB}\right) $ takes into account all the
mirror contributions (qubit-qubit set) as defined in (\protect\ref{GQD3}). A
characteristic time $\protect\tau _{M}$ can be defined when these
correlations are maxima.}
\label{fig-GQD}
\end{figure}

In the figure \ref{fig-GQD}(a) we show the mirror bipartition used in this
work (a similar bipartition has been used for a spin system under the $%
SU\left( 2\right) $ projection \cite{Zanardy}). In the present case, tracing
out (in the lattice) sites different from $\pm s$ it is possible to define a
three-level system. Thus, in order to trace over all non-mirror sites ($\neq
\pm s$ ) we defined the kets 
\begin{eqnarray*}
\arrowvert A\rangle &\leftrightarrow &\arrowvert s\rangle \\
\arrowvert B\rangle &\leftrightarrow &\arrowvert-s\rangle \\
\arrowvert\phi \rangle &\leftrightarrow &\arrowvert s^{\prime }\rangle ,\
s^{\prime }\neq \pm s.
\end{eqnarray*}%
Then, the ket $\arrowvert s_{1},s_{2}\rangle $, representing a state of two
particles, can be written in the form 
\begin{equation}
\arrowvert s_{1},s_{2}\rangle =\arrowvert\alpha \beta \rangle \otimes %
\arrowvert R\rangle ,  \label{GQD22}
\end{equation}%
where $\left\{ \alpha ,\beta \right\} \in \left\{ A,B,\phi \right\} $, and $%
R\ $is the complement, i.e., the set of all non-mirror sites.

Replacing (\ref{GQD22}) in (\ref{rhog}) and tracing over $\arrowvert\phi
\rangle $ we obtain the density matrix $\rho _{AB}$. Thus $\rho _{AB}$ turns
to be a reduced $\left( 9\times 9\right) $ matrix, where the new ordered
basis can be written as:%
\[
\{\arrowvert AA\rangle ,\arrowvert AB\rangle ,\arrowvert A\phi \rangle ,%
\arrowvert BA\rangle ,\arrowvert BB\rangle ,\arrowvert B\phi \rangle ,%
\arrowvert\phi A\rangle ,\arrowvert\phi B\rangle ,\arrowvert\phi \phi
\rangle \}. 
\]

In order to simplify this analysis we now reduce the representation to a 
\textit{qubit-qubit} set, then we do not consider elements of the density
matrix $\rho _{AB}$ with vectors contribution: 
\[
\{\arrowvert A\phi \rangle ,\arrowvert B\phi \rangle ,\arrowvert\phi
A\rangle ,\arrowvert\phi B\rangle ,\arrowvert\phi \phi \rangle \}, 
\]%
i.e., representing the basis for the one-particle state and the empty state.

Therefore final density matrix reduces to a $\left( 4\times 4\right) $
matrix, within this approach we obtain the representation of a qubit-qubit
system. Then, the lower bound of the GQD from Eq.\ref{GQD2}) is reduced to:

\begin{equation}
D_{G}\left( \rho \right) \geq \frac{1}{4}\left( \left\vert \left\vert \vec{x}%
\right\vert \right\vert ^{2}+\left\vert \left\vert T\right\vert \right\vert
^{2}-k_{max}\right) ,  \label{GQD-Q}
\end{equation}%
where $k_{max}$ is the largest eigenvalue of $K=||\vec{x}||^{2}+||T||^{2}$ 
\cite{GQD}. Now, we calculate the total mirror contribution for the GQD
defined as 
\begin{equation}
D_{G}^{T}\left( \rho _{AB}\right) =\sum_{s=1}^{\infty }D_{G}^{\left(
s\right) }\left( \rho _{AB}\right) ,  \label{GQD3}
\end{equation}%
where $D_{G}^{\left( s\right) }\left( \rho _{AB}\right) $ corresponds to (Eq.%
\ref{GQD-Q}) for a fixed value of $s$ ($D_{G}^{\left( s\right) }\left( \rho
_{AB}\right) $ measures the quantum correlation between particles $1$ and $2$
to be confined at sites $\pm s.$). We have plotted $D_{G}^{T}\left( \rho
_{AB}\right) $ as a function of time $t^{\prime }=t_{\Omega }$, for
different values of $r_{D}\equiv \frac{2D}{\Omega /\hbar }$. In figure \ref%
{fig-GQD}(b) the GQD (lower bound given by Eq.\ref{GQD-Q}) is shown for
different values of $r_{D}=0.1,\;0.5,\;1,\;2$.

One important conclusion from this result is that bath-induced correlations
(between the particles) are in fact of quantum nature because $%
D_{G}^{T}\left( \rho _{AB}\right) >0$ for almost all $t>0$ and $r_{D}>0$.
Note that only if $r_{D}=0$ the GQD vanishes at all times. From figure \ref%
{fig-GQD}(b), we can see that the GQD shows a non-monotonic behavior as
function of $r_{D}$ then a characteristic time-scale $\tau _{M}$ can be
defined signing its maximum value; note that as $r_{D}$ decreases $\tau _{M}$
is delayed. In this figure we have not plotted the long-time behavior of
GQD\ because we have numerical computed limitations.

\section{Phase-Space (lattice) Representation}

A important point of view is achieved if we introduce a quasi probability
distribution function (pdf) for the infinite dimensional \textit{discrete}
Hilbert space associated to \textit{two} DQW's. The crucial point in
defining a Wigner function is to assure the completeness of the phase-space
representation \cite{Wigner,Wooters}. Then for this purpose we consider the 
\textit{enlarged} \textit{lattice} of integers ($\mathcal{Z}$) and
semi-integers ($\mathcal{Z}_{2}$). Denoting%
\[
\vec{k}=(k_{1},k_{2}),\ \vec{x}=(x_{1},x_{2}),\ x_{j}\in (\mathcal{Z}\oplus 
\mathcal{Z}_{2}) 
\]%
we define%
\begin{equation}
W_{t}(\vec{k},\vec{x})\!\!\!\ =\!\!\!\!\!\!\!\!\!\!\sum_{x_{1}^{\prime
},x_{2}^{\prime }\in (\mathcal{Z}\oplus \mathcal{Z}_{2})}\!\!\!\!\!\!\!\!\!%
\!\langle x_{1}+x_{1}^{\prime },x_{2}+x_{2}^{\prime }\arrowvert\rho _{t}%
\arrowvert x_{1}-x_{1}^{\prime },x_{2}-x_{2}^{\prime }\rangle \frac{e^{-i2%
\vec{k}\cdot \vec{x}^{\prime }}}{\left( 2\pi \right) ^{2}},  \label{w1}
\end{equation}%
which indeed fulfills 
\[
\sum_{\vec{x}\in (\mathcal{Z}\oplus \mathcal{Z}_{2})}\int \int\limits_{-\pi
}^{\pi }d\vec{k}\ W_{t}(\vec{k},\vec{x})=1 
\]%
Where $\rho _{t}\equiv \rho (t)$ is the two-body density matrix. Note that
when $\rho (t)$ satisfies the exchange of particles symmetry (due to the
particular initial condition we have used) our definition of the Wigner
function fulfills also the invariance under exchange of particles but in the
phase-space; i.e., under the exchange $x_{1}\leftrightarrow x_{2}$ and $%
k_{1}\leftrightarrow k_{2}$ (indicating that $W_{t}(\vec{k},\vec{x})$ has a
reflection symmetry in the planes: $x_{1}-x_{2}=0$ and $k_{1}-k_{2}=0$).

The present definition of $W_{t}(\vec{k},\vec{x})$ can be proved to be
equivalent to the definition using \textit{phase-point operators} in finite
systems \cite{Wooters,Paz,Hinarejos}. We\textit{\ remark} the prescription
that%
\[
\langle \vec{x}\arrowvert\rho _{t}\arrowvert\vec{x}^{\prime }\rangle =0, 
\]%
if some index $x_{j}\in \mathcal{Z}_{2}$ (this is so because Wannier's index
are on $\mathcal{Z}$). Thus, our $W_{t}(\vec{k},\vec{x})\!\!\!$ \ fulfills
the fundamental conditions pointed out by Wigner et al. \cite{Wigner}: $\int
\int\limits_{-\pi }^{\pi }d\vec{k}\ W_{t}(\vec{k},\vec{x})=\langle \vec{x}%
\arrowvert\rho _{t}\arrowvert\vec{x}\rangle \geq 0$ and $\sum_{\vec{x}\in (%
\mathcal{Z}\oplus \mathcal{Z}_{2})}W_{t}(\vec{k},\vec{x})=\langle \vec{k}%
\arrowvert\rho _{t}\arrowvert\vec{k}\rangle \geq 0$. In addition, we noted
that the \textit{enlarged} \textit{lattice} ($\mathcal{Z}\oplus \mathcal{Z}%
_{2}$) is the crucial key for a correct definition of a Wigner function.
From the discrete Fourier transform we can obtain the \textit{inverse }%
relation on the Wannier lattice ($s_{j}\in \mathcal{Z},\forall j=1,2$) as:%
\[
\langle s_{1},s_{2}\arrowvert\rho _{t}\arrowvert s_{1}^{\prime
},s_{2}^{\prime }\rangle =\int \int\limits_{-\pi }^{\pi }d\vec{k}\
W_{t}(k_{1},k_{2},\frac{s_{1}+s_{1}^{\prime }}{2},\frac{s_{2}+s_{2}^{\prime }%
}{2})e^{i\vec{k}\cdot (\vec{s}-\vec{s}^{^{\prime }})}. 
\]%
Note that the inverse relation demands the necessity of a Wigner function
defined on the \textit{enlarged lattice}. After some algebra using solution (%
\ref{rhog}) in definition (\ref{w1}) we get%
%
%
%

\begin{eqnarray}
W_{t}(\vec{k},\vec{x})\!\!\! &=&\!\!\!\frac{e^{-2t_{D}}}{4\pi ^{2}}%
\;(-1)^{2x_{1}+2x_{2}}\!\!\!\!\!\!\!\!\!\!\sum\limits_{\{\alpha ,\beta
,q,n_{2},n_{3},n_{5}\}\in \emph{Z}}\!\!\!\!\!\!\!\!(-1)^{n_{2}+n_{3}+q}
\label{Wigner1} \\
&\times &\!\!\!\!J_{2x_{1}+2\alpha -q}(-2t_{\Omega }\sin
k_{1})J_{2x_{2}+2\beta +q}(-2t_{\Omega }\sin k_{2})  \nonumber \\
&\times
&\!\!\!\!I_{n_{2}}(t_{D})I_{n_{3}}(t_{D})I_{n_{5}}(t_{D})I_{n_{2}+n_{5}-%
\alpha }(t_{D})  \nonumber \\
&\times &\!\!\!\!I_{n_{3}+n_{5}+\beta
}(t_{D})I_{n_{2}+n_{3}+n_{5}-q}(t_{D})e^{iq(k_{1}-k_{2})}.  \nonumber
\end{eqnarray}

\begin{figure}[t]
\includegraphics[width=8.2cm,height=8.5cm]{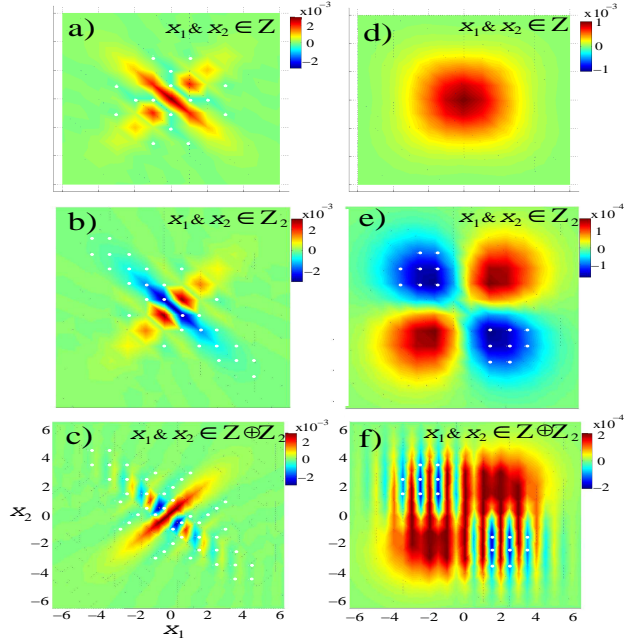}
\caption{(Color online) Wigner function on the enlarged lattice for two
DQW's as a function of $\{x_{1},x_{2}\}\in (\mathcal{Z}$ $\oplus $ $\mathcal{%
Z}_{2})$ for $t_{D}\equiv 2Dt=5$ and two values of $\Delta k$. Due to the
conservation of the pseudo-momentum Wigner's function does only depends on $%
\Delta k$. Domains where $W_{t}$ is negative are shown in dark blue (white
points are put to lighten when starts to be $W_{t}<0$). In (a),(b),(c) $%
\Delta k=0$, and (d),(e),(f) $\Delta k=\protect\pi /3$. }
\label{fiq-Wigner}
\end{figure}

As we commented before this solution is symmetric under exchange of
particles because we have used a localized initial condition. In the case $%
D=0$ we recover the non-dissipative description%
\[
W_{t}(\vec{k},\vec{x})_{D=0}=\frac{(-1)^{2x_{1}+2x_{2}}}{4\pi ^{2}}%
J_{2x_{1}}(-2t_{\Omega }\sin k_{1})J_{2x_{2}}(-2t_{\Omega }\sin k_{2})
\]%
representing two-independent quantum walks, and showing the possibility to
be negative depending on the argument of the Bessel's functions and sites on
the enlarged lattice. Thus, our Wigner function $W_{t}(\vec{k},\vec{x})$ can
be used to detect whether a point in phase-space has a pure quantum
mechanics character or not. In Fig.\ref{fiq-Wigner} we show several
portraits, in particular it is clear to identify regions where $W_{t}(\vec{k}%
,\vec{x})<0$. Therefore we propose to use $W_{t}(\vec{k},\vec{x})$ to
measure the quantum to classical transition. In Fig.\ref{fiq-Wigner}\ we
show patterns for several values of $\Delta k\equiv k_{1}-k_{2}$ and $t_{D}=5
$ on the \textit{enlarged lattice} $\mathcal{Z}\oplus \mathcal{Z}_{2}$. On
the other hand, putting $\Omega =0$ in $W_{t}(\vec{k},\vec{x})$ simplifies
its analytical expression and shows also that there are domains where it is
negative. This proves the quantum mechanics character of the bath-induced
correlations between the two DQW, even in the large dissipative regime $%
\frac{2D}{\Omega /\hbar }\gg 1$. 
Note that the behavior of the \textit{Negative Volume} of Wigner's function
in phase-space can enlighten the understanding of the quantum character of
the system behavior, as well as supporting the characterization of the
typical \textit{time-scale} for the coherence between particles, work along
this line is in progress.

\section{Discussions}

We have analyzed two \textit{free} spinless initially uncorrelated particles
(in the lattice) in interaction with a common boson thermal bath $\mathcal{B}
$. Even when the QME is a second order approach, the Markov approximation is
enough to show bath-induced correlation among free particles. We have solved
analytically the QME showing that if $D\neq 0$, we get $\rho (t)\neq \rho
_{1}(t)\otimes \rho _{2}(t),\forall t>0$, i.e., the time evolution is not a
direct product of two independent particles if the temperature of the bath
is non null. For $D=0$ the probability profile $P_{s_{1},s_{2}}(t)\equiv
\langle s_{1},s_{2}|\rho (t)|s_{1},s_{2}\rangle $ is ballistic and starts to
be modified by the presence of dissipation $D>0$, showing a \textit{X-form}
pattern. In the case of large dissipation, $r_{D}\equiv \frac{2D}{\Omega
/\hbar }\gg 1$, this structure is accentuated and additional interferences
are observed. Several correlations measures: $\mathcal{T}_{(1,2)}$, $%
\mathcal{C}_{(1,2)}$, Purity and Relative Entropy of Coherence $C_{RE}$ have
been analyzed showing a degree of coherence between particles, these
correlations are induced by the common bath $\mathcal{B}$ despite the
presence of dissipation for temperature $T\neq 0$. $\mathcal{P}_{Q}$, Mirror
Correlation $\mathcal{T}_{(1,2)}$ and GQD have been used to show the
existence of a \textit{time-scale} when the quantum correlations reach a
maximum. All these measured of correlations have also been indirectly
supported by an independent analysis using a Wigner function defined on the
enlarged lattice of integers and semi-integers ($\mathcal{Z}\oplus \mathcal{Z%
}_{2}$); showing that this function is negative in some domains of
phase-space. Thus we propose to use the total \textit{negative volume} of
the Wigner function in phase-space to characterize the quantum to classical
transition in this type of many-body system, work along this line is in
progress.


\textit{Acknowledgments.} M.O.C and M.N. gratefully acknowledge support
received from CONICET, grant PIP 112-261501-00216 CO, Argentina.

\section{Appendix A. Semigroup for two distinguishable quantum random walk}

Starting from the total Hamiltonian $H_{\mathcal{T}}$ 
and eliminating the bath variables (in the Markov approximation), the
Kossakowski-Lindblad infinitesimal generator \cite{Alicki} can be written in
the form 
\begin{equation}
\mathcal{KL}\left[ \bullet \right] =-\frac{i}{\hbar }\left[ H_{eff},\bullet %
\right] +F\left[ \bullet \right] -\frac{1}{2}\left\{ F^{\ast }\left[ \mathbf{%
1}\right] ,\bullet \right\} _{+},  \label{KL1}
\end{equation}%
where $H_{eff}$ is the effective Hamiltonian, $\frac{1}{2}F^{\ast }\left[ 
\mathbf{1}\right] $ can be regarded as a dissipative operator, and $F\left[
\bullet \right] $ the fluctuating superoperator ($F^{\ast }\left[ \mathbf{%
\bullet }\right] $ is the dual operator of $F\left[ \mathbf{\bullet }\right] 
$, and $\left\{ \bullet ,\bullet \right\} _{+}$ the anticonmutator). Using a
separable initial condition for the total density matrix $\rho _{T}(0)=\rho
(0)\otimes \rho _{\mathcal{B}}^{e}$, and working in a second order
approximation we can write ($\rho _{\mathcal{B}}^{e}$ is the thermal density
matrix of the bath at temperature $T$)%
\begin{eqnarray*}
H_{eff} &=&H_{\mathcal{S}}-i\frac{1}{2\hbar }\int_{0}^{\infty }d\tau \ \text{%
Tr}_{\mathcal{B}}\left( \left[ H_{\mathcal{SB}},H_{\mathcal{SB}}\left( -\tau
\right) \right] \rho _{\mathcal{B}}^{e}\right) , \\
F\left[ \rho (t)\right] &=&\left( \frac{1}{\hbar }\right)
^{2}\int_{0}^{\infty }d\tau \ \text{Tr}_{\mathcal{B}}\left[ H_{\mathcal{SB}%
}\rho (t)\otimes \rho _{\mathcal{B}}^{e}H_{\mathcal{SB}}\left( -\tau \right)
\right. \\
&&+\left. H_{\mathcal{SB}}\left( -\tau \right) \rho (t)\otimes \rho _{%
\mathcal{B}}^{e}H_{\mathcal{SB}}\right] ,
\end{eqnarray*}%
where $H_{\mathcal{SB}}\left( -\tau \right) =e^{-i\tau \left( H_{\mathcal{S}%
}+H_{\mathcal{B}}\right) /\hbar }\ H_{\mathcal{SB}}\ e^{i\tau \left( H_{%
\mathcal{S}}+H_{\mathcal{B}}\right) /\hbar }$ \cite{Alicki,manuel}. Noting
that $a_{12}$ and $a_{12}^{\dag }$ are constant in time, $\left[
a_{12},a_{12}^{\dag }\right] =0$ and using the full expressions of $H_{%
\mathcal{S}},H_{\mathcal{B}}$ and $H_{\mathcal{SB}}$, after some algebra we
can write%
\begin{equation}
F\left[ \bullet \right] =\frac{\pi 4\Gamma ^{2}}{2\hbar /k_{B}T}\left[
a_{12}\bullet a_{12}^{\dag }+a_{12}^{\dag }\bullet a_{12}\right] ,
\label{CP1}
\end{equation}%
where $\pi \Gamma ^{2}k_{B}T/\hbar $ is a dissipative constant (here we have
used the Ohmic approximation for the \textit{strength function }$g\left(
\omega \right) $ of the bath, i.e., $g(\omega )=\sum_{k}\left\vert
v_{k}\right\vert ^{2}\delta \left( \omega -\omega _{k}\right) \propto \omega 
$, if $0<\omega <\tilde{\omega}_{c}$). In a similar way the effective
Hamiltonian can be calculated given%
\[
H_{eff}=H_{\mathcal{S}}-\omega _{c}\hbar \ a_{12}a_{12}^{\dag }, 
\]%
here $\omega _{c}\equiv 2\tilde{\omega}_{c}\Gamma ^{2}$ is an upper bound
frequency.

Using these expressions we can write down the QME (\ref{QME}) in the form%
\[
\dot{\rho}=L\left[ \rho \right] ,\quad L\left[ \bullet \right] \equiv -\frac{%
i}{\hbar }\left[ H_{eff},\bullet \right] +F\left[ \bullet \right] -\frac{1}{2%
}\left\{ F\left[ \mathbf{1}\right] ,\bullet \right\} _{+}, 
\]%
then the solution can be written in the formal form: 
\[
\rho \left( t\right) =\sum_{m=0}^{\infty
}\int_{0}^{t}dt_{m}\int_{0}^{t_{m}}dt_{m-1}\cdots
\int_{0}^{t_{2}}dt_{1}\left\{ \mathcal{L}_{0}\left( t-t_{m}\right) F\left[
\bullet \right] \mathcal{L}_{0}\left( t_{m}-t_{m-1}\right) \cdots F\left[
\bullet \right] \mathcal{L}_{0}\left( t_{1}\right) \right\} \rho \left(
0\right) , 
\]%
where it is evidenced that the system is exposed to a succession of quantum
jumps associated to the superoperator $F\left[ \bullet \right] $, and
intercalating a smooth evolution characterized by%
\[
\mathcal{L}_{0}\left( t\right) \rho =\exp \left\{ \left( -\frac{i}{\hbar }%
\left[ H_{eff},\bullet \right] -\frac{1}{2}\left\{ F\left[ \mathbf{1}\right]
,\bullet \right\} _{+}\right) t\right\} \rho . 
\]

This picture allows to generalize the description of a QDW\ into a
non-Markovian evolution using the CTRW approach \cite{Montroll,libro}. See
also a related contribution, in the present issue, for describing completely
positive quantum maps in the context of the \textit{Renewal theory} \cite%
{caceres2017}.


\begin{thebibliography}{99}
\bibitem{Reichl} L.E. Reichl, \textit{A Modern Course in Statistical
Mechanics} (E. Arnold, Publ. LTD), Uni. of Texas Press, (1980).

\bibitem{holanda} A.E. Allahverdayn et al., Phys. Rep., 525, 1-166, (2013).

\bibitem{Alicki} R. Alicki and K. Lendi, \textit{Quantum Dynamical
Semigroups and Applications} (Spronger Verlag), Lecture Notes in Physics,
V.286, Berlin,(1987).

\bibitem{nielsen} M. Nielsen and I. Chuang, \textit{Quantum Computation and
Quantum Information} (Cambridge University Press, Cambridge, (2000); and
references therein.

\bibitem{Kolobov} M.I. Kolobov and C. Fabre, Phys. Rev. Lett. \textbf{85},
3789, (2000); C.M. Caves, Phys. Rev. D, \textbf{23}, 1693, (1981).

\bibitem{Schlosshauer} M. Schlosshauer, Decoherence and the
Quantum-to-Classical Transition (Springer, Berlin, 2007).

\bibitem{Kendon} V. Kendon, Math. Struct. in Comp. Sci \textbf{17}, 1169
(2006).

\bibitem{Braun} D. Braun, Phys Rev. Lett. 89, 277901, (2002).

\bibitem{Finite} K. Ann and G. Jaeger, Phys Rev. A, 76, 044101 (207); T. Yu
and J.H. Eberly. Phys Rev. Lett. 97, 140403 (2006); T. Yu and J.H. Eberly,
Phys Rev. B 68, 165322 (2003); D.P.S. McCutcheon, A. Nazir, S. Bose, and
A.J. Fisher, Phys. Rev. A, 80, 022337 (2009).

\bibitem{CommBath} T. Yu and J.H. Eberly, Phys Rev. B 66, 193306 (2002);
M.J. Storcz,.U. Hartmann, S. Kohler, and F.K. Wilhelm, Phys. Rev. B 72,
235321 (2005).

\bibitem{ChemPhysLett} M. Thorwart, J. Eckel, J.H. Reina, P. Nalbach, S.
Weiss, Chem. Phys. Lett. 478, 234 (2009).

\bibitem{aharanov} Y. Aharonov et al., Phys. Rev. A \textbf{48}, 1687 (1993).

\bibitem{vK95} N.G. van Kampen; J. Stat. Phys. \textbf{78}, 299 (1995).

\bibitem{blumen2} O. M\"{u}lken and A. Blumen, Phys. Rev. E, \textbf{71},
036128 (2005).

\bibitem{manuel} M.O. C\'{a}ceres and M. Nizama, J. Phys. A: Math. Theor. 
\textbf{43} 455306 (2010).

\bibitem{physA-1} M. Nizama and M.O. C\'{a}ceres, Physica A \textbf{400}, 31
(2014)

\bibitem{physA-2} M. Nizama and M.O. C\'{a}ceres, Physica A \textbf{392},
6155, (2013).

\bibitem{zahringer} F. Z\"{a}hringer et al., Phys. Rev. Lett. \textbf{104},
100503 (2010)

\bibitem{schreiber} A. Schreiber et al., Phys. Rev. Lett. \textbf{106},
180403 (2011).

\bibitem{broome} M.A. Broome et al., Phys. Rev. Lett. \textbf{104}, 153602
(2010).

\bibitem{schmitz} H. Schmitz et al., Phys. Rev. Lett. \textbf{103}, 090504
(2009).

\bibitem{karski} M. Karski et al., Science \textbf{325}, 174 (2009).

\bibitem{kempe} J. Kempe, Contemp. Phys. \textbf{44}, 307 (2003).

\bibitem{mm} M. Nizama and M.O. C\'{a}ceres, J. Phys. A: Math. Theor. 
\textbf{45}, 335303 (2012).

\bibitem{Montroll} E.W. Montroll and G.H. Weiss, J. Math. Phys. \textbf{6},
167, (1965).

\bibitem{libro} M.O. C\'{a}ceres \textquotedblleft \textit{Non-equilibrium
Statistical Physics with Application to Disordered Systems}%
\textquotedblright , ISBN 978-3-319-51552-6, Springer (2017).

\bibitem{caceres2017} M.O. C\'{a}ceres, Euro Physics Journal B (2017) 90: 74
DOI: 10.1140/epjb/e2017-80009-8.

\bibitem{Two-QW} This free Hamiltonian was also presented in the context of
quantum walk, see: C. Benedetti, F. Buscemi, \& P. Bordone in: Phys. Rev. A, 
\textbf{85}, 42314 (2012).

\bibitem{particles} To prove the invariance under the exchange of particles: 
$\left\{ s_{1}\leftrightarrow s_{2},s_{1}^{\prime }\leftrightarrow
s_{2}^{\prime }\right\} $, note that in Eq.(\ref{rhog}) $n_{j}$ are mute
variables therefore we can use the change of variables $n_{1}\leftrightarrow
n_{6},n_{2}\leftrightarrow n_{3}$ and finally $n_{4}\leftrightarrow
-n_{4},n_{5}\leftrightarrow -n_{5}$ to check this symmetry.

\bibitem{Plenio2014} T. Baumgratz, M. Cramer, and M. B. Plenio, Phys. Rev.
Lett. \textbf{113}, 140401 (2014).

\bibitem{mauro} Mauro B. Pozzobom and Jonas Maziero (arxiv.org)
quant-ph/1605.04746 (2016).


\bibitem{GQD} B. Dakic, V. Vedral, and C. Brukner, Phys. Rev. Lett. \textbf{%
105}, 190502 (2010).

\bibitem{lowerGQD} A. S. M. Hassan, B. Lari, and P. S. Joag, Phys. Rev. A. 
\textbf{85}, 0243202 (2012).

\bibitem{3GQD} S. Rana, and P. Parashar, arXiv:1201.5969v2[quant-ph] 13 Feb
2012.

\bibitem{Rau} Sal Vinjanampathy and A.R.P. Rau, J. Phys. A 45, 095303 (2012).

\bibitem{Discord} L. Henderson and V. Vedral, J. Phys. A \textbf{34}, 6899
(2001); H. Ollivier and W.H. Zurek, Phys. Rev. Lett. \textbf{88}, 017901
(2001).

\bibitem{Zanardy} P. Zanardi et al. J. Phys. A Math. Gen. \textbf{35}, 7947
(2002).

\bibitem{Wigner} M. Hillery et al., Phys. Rep. \textbf{106}, 121 (1984).

\bibitem{Wooters} W.K. Wooters, Ann. Phys. (N.Y.) 176, 1, (1987).

\bibitem{Paz} C. Miquel et al., Phys. Rev. A, 65, 62309, (02).

\bibitem{Hinarejos} M. Hinarejos et al., N. J. Phys. \textbf{14}, 103009,
(2012).
\end{thebibliography}
\end{document}